\documentclass[doublecol]{epl2} 
\pdfoutput=1
\usepackage{subfigure}
\usepackage{amssymb}
\usepackage{amsfonts}
\usepackage{amsmath}
\usepackage{epsfig}
\usepackage{graphicx}
\usepackage{color}

\newcommand{\lrangle}[1]{\langle{#1}\rangle}

\newcommand{\Oc}[1]{\mathcal{O}({#1})}

\title{Pair quenched mean-field theory for the susceptible-infected-susceptible
model on complex networks}
\shorttitle{Pair QMF theory for the SIS model on complex networks} 
\author{Ang\'elica S. Mata \and Silvio C. Ferreira}
\shortauthor{Mata and Ferreira}
\institute{                    
  Departamento de F\'{\i}sica - Universidade Federal de Vi\c{c}osa, 36571-000, Vi\c{c}osa,
Minas Gerais, Brazil
}

\abstract{
We present a quenched mean-field (QMF) theory for the dynamics of the
susceptible-infected-susceptible (SIS) epidemic model on complex networks where
dynamical correlations between connected vertices are taken into account
by means of a pair approximation. 
We present analytical expressions of the epidemic thresholds in the star and
wheel graphs  and in random regular networks. For random networks with a power
law degree distribution, the thresholds are numerically determined via an
eigenvalue problem. The pair and one-vertex QMF theories yield the same scaling 
for the thresholds as functions of the network size. However, comparisons
with quasi-stationary  simulations of the SIS dynamics on large networks show
that the former is quantitatively much more accurate than the latter. Our
results demonstrate the central role played by dynamical correlations on the
epidemic spreading and introduce an efficient way to theoretically access
the thresholds of very large networks that can be extended to 
dynamical processes in general.}

\begin{document}

\maketitle


The propagation of an epidemic is an ultimately important issue 
in a broad collection of systems ranging from  the disease
dissemination within a population~\cite{anderson92,keeling} to the virus spreading
throughout a computer network~\cite{alexei,Romu,Lloyd18052001} among a plenty of
examples~\cite{Newman10}. The description of the epidemic dynamics has drawn
the attention of biologist and mathematicians since a long
time~\cite{anderson92} and, more recently, of the statistical physics~\cite{alexei,Romu} and
computer science communities~\cite{Wang03,Ganesh05}. The simplest epidemic model is the
susceptible-infected-susceptible (SIS) model where individuals in a population can
be only in one of two states: infected or susceptible (healthy). Infected
individuals become spontaneously healthy at rate 
$\delta=1$ (this choice fixes the time scale) while the susceptible ones are infected at
rate $\lambda n$, where $n$ is the number of infected contacts of the
individual~\cite{anderson92}. The SIS dynamics exhibits a phase transition
between a disease-free (absorbing) state and an active stationary phase where a fraction of
the population is infected. These regimes are separated by an epidemic threshold
$\lambda_c$, the core issue of many recent works about epidemic
spreading~\cite{Romu,Wang03,Ganesh05,Chatterjee09,Chakrabarti,Durrett10,
newman02,Castellano10,Castellano12,VanMieghem09,Ferreira12,VanMieghem12,Gomez10,Goltsev12}.

The former investigations of the epidemic models relied on the hypothesis of a
homogeneous mixing of the population~\cite{anderson92},  neglecting the highly
heterogeneous structure of the contact network inherent to real
systems~\cite{barabasi02}. Indeed, many bio, socio and technological systems are
characterized by heavy tailed distributions of the number of contacts $k$ of an
individual (the vertex degree, in the network language), for which homogeneity 
hypothesis is severely violated~\cite{barabasi02,Newman10,barratbook}.  Complex
networks are, in fact, a framework where the  heterogeneity of the contacts can be
naturally afforded~\cite{barabasi02}.

The heterogeneous mean-field (HMF) theory is a benchmark for
dynamical process running on the top of complex networks~\cite{barratbook,dorogovtsev07}. 
In a HMF theory,  dynamical
quantities, as the density of infected individuals in the SIS model, depend only
of the vertex degree and do not of their specific
location in the network. The HMF epidemic threshold for the SIS dynamics 
in undirected and uncorrelated networks is given by~\cite{Romu}
\begin{equation}
\lambda_c^{hmf}={\lrangle{k}}/{\lrangle{k^2}},
 \label{eq:lbchmf}
\end{equation} 
where $\lrangle{k^n}=\sum_kk^nP(k)$ and $P(k)$ is the degree distribution
defined as the probability that a randomly chosen vertex has degree $k$.
Equation~(\ref{eq:lbchmf}) has strong
implications since several real networks have a power law degree distribution
$P(k)\sim k^{-\gamma}$ with exponent in the range
$2<\gamma<3$~\cite{barabasi02}. For these distributions, the second moment $\lrangle{k^2}$ diverges in the
limit of infinite sizes implying a  vanishing threshold for the
SIS model.

An alternative approach, called of quenched mean-field (QMF)
theory~\cite{Castellano10}, has been developed in parallel to the HMF
theory~\cite{Wang03,Chakrabarti,Durrett10,Ganesh05,VanMieghem12,Goltsev12,Gomez10}. The QMF theory
explicitly takes into account the actual connectivity of the network through
its adjacency matrix. 
The epidemic threshold of the SIS model in a QMF approach
is~\cite{Castellano10}
\begin{equation}
\lambda_c^{qmf}={1}/{\Lambda_m},
 \label{eq:lbcqmf}
\end{equation}
where $\Lambda_m$ is the largest eigenvalue of the adjacency matrix. 
The noteworthy differences
and similarities between HMF and QMF theories were realized in
Ref.~\cite{Castellano10}.
The central point is that $\Lambda_m$ diverges for
increasing networks with power law degree distributions
even when $\lrangle{k^2}$ remains finite~\cite{chung03}.
Both theories predict vanishing thresholds for $\gamma<3$ despite of different
scaling for $5/2<\gamma<3$. However, while HMF predicts a finite threshold for
networks with $\gamma>3$, QMF still predicts a
vanishing threshold~\cite{Castellano10}.
{Very recently, semi-analytic methods including local and long-range
dynamical fluctuations~\cite{Lee2013,boguna2013nature} have been proposed
as alternatives to HMF and QMF theories.}

A numerical investigation of the thresholds of the SIS model on several networks
was recently done in Ref.~\cite{Ferreira12}. It was shown that the  QMF theory
is an improvement of HMF theory but it is still an approximation. In fact, the
assumption where the probability that  a vertex is infected does not depend of
the states of its neighbors is used in both approaches. This approximation
neglects all dynamical correlations between vertices, which possibly contribute
to the threshold value. Interestingly, the thresholds of some models with
absorbing configurations taking place in highly heterogeneous networks 
are surprisingly well described by a simple homogeneous  pair
approximation~\cite{FFCR11,Juhasz12,sander_phase_2013}, the simplest mean-field 
theory that considers dynamical correlations between vertices.

In the present work, we develop an extension of the QMF theory for the SIS
model, the pair QMF theory, using  a heterogeneous pair mean-field
approximation. Analogously to the one-vertex QMF theory, our perturbative 
theory includes the actual connectivity of the network through its adjacency
matrix. 
Analytical expressions are presented for the random regular networks, star and
wheel graphs. We also investigated large random networks with a power law degree
distribution. The thresholds obtained in pair and one-vertex QMF theories have
the same scaling with the system size but the pair QMF theory is quantitatively 
much more accurate than the one-vertex theory when compared with simulations.

To develop the pair QMF theory, we introduce the following notation: $[A_i]$ 
is probability that the vertex $i$ is in
the state $A$; $[A_i,B_j]$ is probability that the vertices $i$ and $j$ are in
states $A$ and $B$, respectively; $[A_i,B_j,C_k]$ is the extension to three
vertices; and so forth. The  infected state is represented by $1$ and the
susceptible one by $0$. We introduce the variables $\rho_i=[1_i]$ and,
consequently, $[0_i]=1-\rho_i$, $\psi_{ij}=[1_i,1_j]$, $\omega_{ij}=[0_i,0_j]$,
$\phi_{ij}=[0_i,1_j]$, and $\bar{\phi}_{ij}=[1_i,0_j]$. Obviously we have that
$\psi_{ij}=\psi_{ji}$, $\omega_{ij}=\omega_{ji}$, and
$\phi_{ij}=\bar{\phi}_{ji}$.
The following relations hold for any pair of vertices
\begin{eqnarray}
\psi_{ij}+\phi_{ij}=\rho_j, ~~~
\psi_{ij}+\bar{\phi}_{ij}=\rho_i \nonumber \\ 
\omega_{ij}+\phi_{ij}=1-\rho_i, ~~~
\omega_{ij}+\bar{\phi}_{ij}=1-\rho_j.\label{eq:prob4}
\end{eqnarray}

The dynamical equation for the vertex $i$ is 
\begin{equation}
 \frac{d\rho_i}{dt} = -\rho_i+\lambda\sum_j\phi_{ij}A_{ij},
\label{eq:rho_iPair}
\end{equation}
where $A_{ij}$ is the adjacency matrix that, for unweighted and 
undirected networks, assumes $A_{ij}=A_{ji}=1$ if vertices $i$ 
and $j$ are connected and $A_{ij}=0$, otherwise. 
Using a one-vertex approximation where the joint probability $\phi_{ij}$
is factorized as $\phi_{ij}\approx (1-\rho_i)\rho_j $,
one obtains
\begin{equation}
 \frac{d\rho_i}{dt} = -\rho_i+\lambda(1-\rho_i)\sum_{j=1}^NA_{ij}\rho_j.
\label{eq:rho_i1}
\end{equation}
Performing a linear stability analysis around the trivial fixed
point $\rho_i=0$, one has  
$\frac{d\rho_i}{dt} = \sum_j L_{ij}\rho_j,$
where the Jacobian matrix is
$L_{ij}=-\delta_{ij}+\lambda A_{ij}$, $\delta_{ij}$ being the Kronecker delta symbol. 
The transition point is defined when the absorbing 
phase becomes unstable or, equivalently, when the largest eigenvalue  
of $L_{ij}$ is null~\cite{Hilborn}.  
So, one directly finds the threshold given by Eq.~(\ref{eq:lbcqmf}).

The dynamical equation for $\phi_{ij}$, 
considering a pair of connected vertices $(i,j)$, is given by
\begin{equation}
 \frac{d\phi_{ij}}{dt}=-\phi_{ij}-\lambda\phi_{ij}+\psi_{ij}
+\lambda\sum_{\substack{l\in\mathcal{N}(j) \\ l\ne i}}             
 [0_i,0_j,1_l]
-\lambda\sum_{\substack{l\in\mathcal{N}(i) \\ l\ne j}}
[1_l,0_i,1_j].
\label{eq:phi1}
\end{equation}
where $\mathcal{N}(i)$ represents the neighborhood of the vertex $i$.
The first three terms represents the reactions involving only the pair $(i,j)$,
that can create or destroy the configuration $[0_i,1_j]$\footnote{
{Spontaneous annihilation events $[0_i,1_j]\rightarrow[0_i,0_j]$ 
and $[1_i,1_j]\rightarrow[0_i,1_j]$ and the creation in vertex $i$ due to 
$j$,  $[0_i,1_j]\rightarrow[1_i,1_j]$.} 
}
while the third and fourth terms represent changes due to the interaction with the
neighbors of $i$ and $j$, 
respectively, excluding the link $(i,j)$
\footnote{
{Creation events in $i$ or $j$ due another vertex $l$, represented by 
transitions $[1_l,0_i,1_j]\rightarrow [1_l,1_i,1_j]$ and 
$[0_i,0_j,1_l]\rightarrow [0_i,1_j,1_l]$, respectively, can also
destroy/create a configuration $[0_i,1_j]$.} }.

Equations~(\ref{eq:rho_iPair}) and (\ref{eq:phi1}) cannot be solved due to the
triplets. However, the dynamical equations for triplets will depend on
quadruplets, and so
forth. 
So, the hierarchy of clusters must be broken in some point to obtain an
approximated solution. 
In the present work, we apply the standard 
pair-approximation~\cite{Avraham92,henkel2008non}
\begin{equation}
 [A_i,B_j,C_l]\approx \frac{[A_i,B_j][B_j,C_l]}{[B_j]}
\label{eq:clu}
\end{equation}
and the adjacency matrix 
in equation~(\ref{eq:phi1}) to obtain
\begin{equation}
\begin{array}{lll}
 \frac{d\phi_{ij}}{dt} & = & -(1+\lambda)\phi_{ij}+\psi_{ij}
+\lambda\sum_{l} \frac{\omega_{ij}\phi_{jl}}{1-\rho_j}(A_{jl}-\delta_{il})\\ & &
-\lambda\sum_{l} \frac{\phi_{ij}\bar{\phi}_{li}}{1-\rho_i}(A_{il}-\delta_{lj}) .
\end{array}
\label{eq:phi2}
\end{equation}
We now perform a linear stability analysis of Eq. (\ref{eq:phi2}) around the  
fixed point $\rho_i =\phi_{ij}=\psi_{ij}=0$ 
to find
\begin{equation}
 \frac{d\phi_{ij}}{dt}=-(1+\lambda)\phi_{ij}+\psi_{ij} 
+\lambda\sum_{l} \phi_{jl}(A_{jl}-\delta_{il}).
\label{eq:phi_lin}
\end{equation}
Performing a quasi-static approximation for $t\rightarrow\infty$, 
$d\rho_i/dt\approx 0$ and 
$d\phi_{ij}/dt\approx 0$, in Eqs.~(\ref{eq:rho_iPair}) and
(\ref{eq:phi_lin}), respectively, one finds 
\begin{equation}
\phi_{ij}\approx \frac{(2+\lambda)\rho_j-\lambda\rho_i}{2+2\lambda},
\label{eq:phi_qstat}
\end{equation}
where the relations given by Eq.~(\ref{eq:prob4}) were 
used to eliminate other dynamical variables.
Substituting Eq.~(\ref{eq:phi_qstat})  in (\ref{eq:rho_iPair}) one finds the 
Jacobian matrix
\begin{equation}
 L_{ij} = -\left(1+\frac{\lambda^2 k_i}{2\lambda+2}\right)\delta_{ij}
+\frac{\lambda(2+\lambda)}{2\lambda+2}A_{ij}.
\label{eq:Lij}
\end{equation}
Again, the critical point is obtained when the largest eigenvalue of
$L_{ij}$ is null. Equation~(\ref{eq:Lij}) is the central result of our work.
Even though we do not present a closed expression for the threshold in an arbitrary
network, we have obtained analytical solutions of transition points for simple
networks. These solutions are very important to test the consistency of the
results and to unveil the basic mechanisms that sustain an epidemic
phase~\cite{Castellano10,Castellano12}. For the general case, the critical point can
be obtained  solving Eq.~(\ref{eq:Lij}) numerically.

An approach similar to ours was developed in Ref.~\cite{Cator12}, where a
different approximation was used to split the joint probabilities in cluster
approximation: $[A_iB_jC_k]\approx [A_iB_j][C_k]$ instead of the standard pair
approximation, Eq.~(\ref{eq:clu}),  which has been generally accepted in the
nonequilibrium statistical community as the most reliable
approach~\cite{henkel2008non}. 

\begin{figure}[ht]
\centering
\includegraphics[width=8cm]{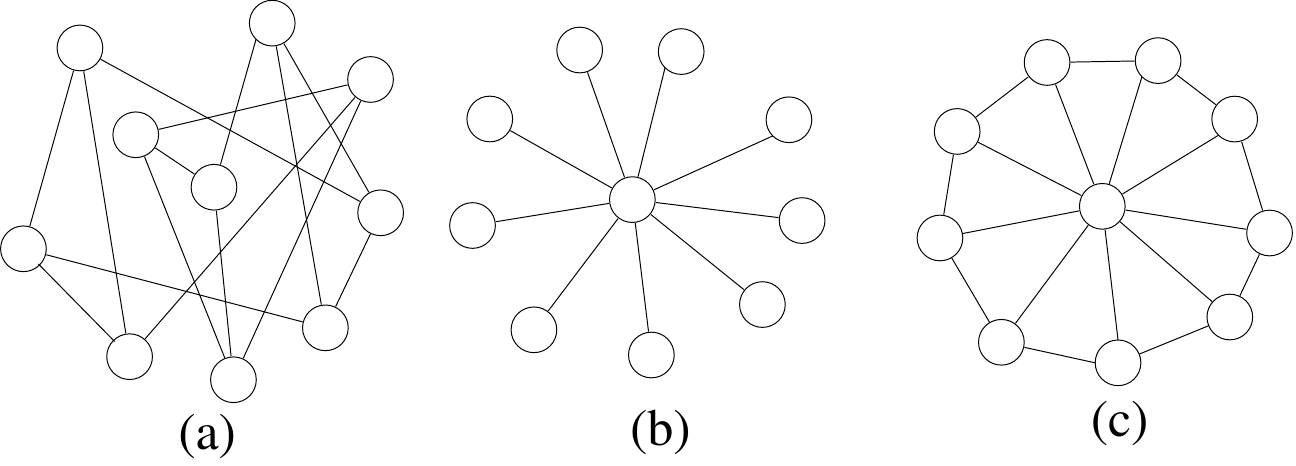}
\caption{Simple graphs used to study SIS dynamics under pair
QMF theory: (a) Random regular networks, (b) star and (c) wheel graphs.}
\label{fig:simple}
\end{figure}

Let us start the investigation of Eq.~(\ref{eq:Lij}) for a simple homogeneous 
network, the random regular network (RRN) illustrated in Fig.~\ref{fig:simple}(a). 
In this network, 
all $N$ vertices have the same connectivity $m$, 
$P(k)=\delta_{km}$, while the connections are done at random, avoiding both self and 
multiple connections. The largest eigenvalue of
$L_{ij}$ is given by
\begin{equation}
 \Upsilon_m = -\left[1+\frac{\lambda^2 m}{2\lambda+2}\right]
+\frac{\lambda(2+\lambda)}{2\lambda+2}m,
\end{equation}
where we used the largest eigenvalue of the $A_{ij}$ given by
$\Lambda_m=m$~\cite{Ferreira12}.
The threshold is then obtained as 
\begin{equation}
 \lambda_c^{pqmf} = \frac{1}{m-1},
\end{equation}
that corresponds to the threshold  of a simple homogeneous pair
approximation~\cite{Ferreira12}. Actually, simulations of the SIS model on RRNs reported
in Ref.~\cite{Ferreira12} showed that the thresholds are, in fact, much closer
to the homogeneous pair approximation than to the standard QMF theory, 
a fact captured by the pair QMF  theory.

\begin{figure}[ht]
 \centering
 \includegraphics[width=8cm]{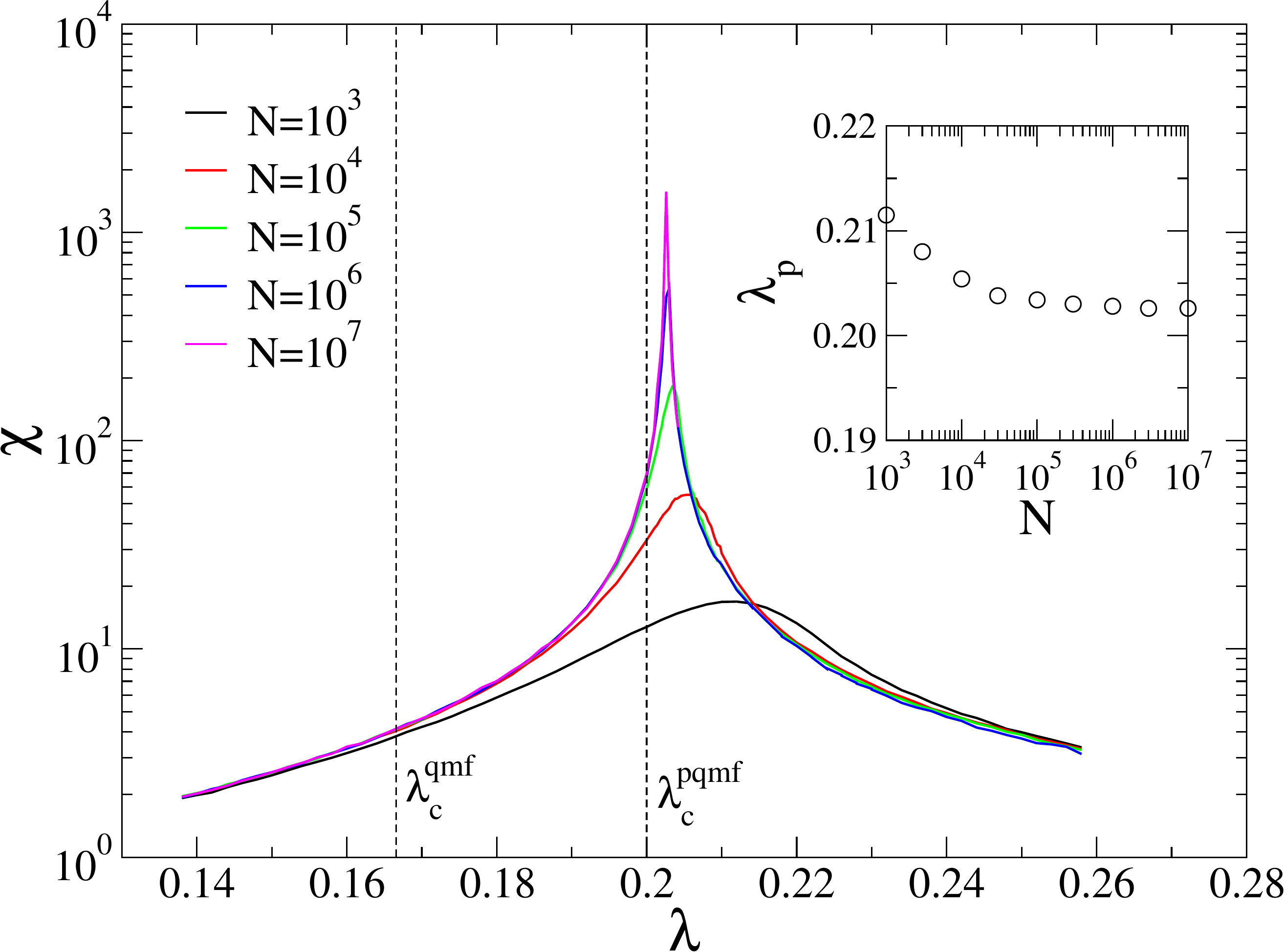}
 \caption{Susceptibility against infection rate for SIS model on RRNs 
with fixed degree $m=6$. Inset shows the positions 
of the peaks of the susceptibility against network size.}
 \label{fig:rrn}
\end{figure}

We performed simulations of the SIS model using the quasi-stationary method
for dynamical processes with absorbing states suitably adapted for 
networks~\cite{FFPS11}. Details of the implementation 
are given in Ref.~\cite{Ferreira12}. The thresholds in finite 
networks can be estimated using the peaks  of the susceptibility $\chi$ defined,
in terms of the density of occupied vertices, as~\cite{Ferreira12}
\begin{equation}
 \chi =N\frac{\lrangle{\rho^2}-\lrangle{\rho}^2}{\lrangle{\rho}}.
\end{equation}
Figure~\ref{fig:rrn} shows the susceptibility against infection rate for $m=6$ exhibiting
a sharp peak that asymptotically converges to a position 
$\lambda_c\simeq0.2026$ that is only 1\% above the theoretical value predicted
by the pair QMF theory, $\lambda_c^{pqmf}=0.2$. So,  the pair QMF theory
improves a lot the estimate of the thresholds in relation to the simple QMF
theory $\lambda_c^{qmf}=0.166$, which errs approximately 22\% the peak position
for $m=6$. Moreover, the larger the vertex degree $m$ the better the pair
approximation. For $m=3$, the peak converges to $\lambda_p=0.5421$ that is
relatively much farther from the pair QMF threshold, $\lambda_c^{pqmf}=0.5$,
than in the case $m=6$. The better accuracy of the pair QMF theory for larger
$m$ is intuitive since  the average distance among vertices decreases as the
average degree increases making the mean-field premise a more credible
hypothesis.

As an example of simple heterogeneous network, we consider star graph 
defined as a hub, $i=0$, connected to $N$ leaves,
$i=1,\ldots,N$ , of degree $k_i=1$, 
as shown in Fig.~\ref{fig:simple}(b). This structure plays a
central role in the SIS dynamics 
since the epidemic activity may be sustained by the sub-graph composed  by
the most connected vertex and its neighborhood for networks with
degree exponent $\gamma>2.5$~\cite{Castellano12}. The adjacency matrix of a star
graph is  $A_{0i}=A_{i0}=1$ for $i=1\ldots N$ and $A_{ij}=0$, otherwise.
Therefore, the elements of the Jacobian $L_{ij}$  are given by
\begin{equation}
L_{ii} = -1-\frac{\lambda^2}{2\lambda+2}[(N-1)\delta_{i0}+1]
\label{eq:star0}
\end{equation}
for the diagonal and 
\begin{equation}
 L_{ij}= \frac{\lambda(2+\lambda)}{(2\lambda+2)}(\delta_{0i}+\delta_{0j})
\end{equation}
for $i\ne j$. 
Thus, the eigenvalue equations for $L_{ij}$ are
\begin{equation}
 -\left(1+\frac{\lambda^2N}{2\lambda+2}\right)v_{0}+
\frac{\lambda(2+\lambda)}{2\lambda+2}\sum_{j=1}^Nv_j=\Upsilon v_{0}
\label{eq:star1}
\end{equation}
and 
\begin{equation}
 \left[\frac{\lambda(2+\lambda)}{2\lambda+2}\right]v_{0}
-\left(1+\frac{\lambda^2}{2\lambda+2}\right)v_i=\Upsilon v_i,~~~i=1\ldots N.
\label{eq:star2}
\end{equation}
Isolating $v_i$ in Eq.~(\ref{eq:star2}) and substituting it
in Eq.~(\ref{eq:star1}), two different eigenvalues are obtained. Taking
the largest one as zero, one obtains  the threshold
\begin{equation}
 \lambda_c = \frac{\sqrt{2N-1}+1}{N-1}\simeq \sqrt{\frac{2}{N}},
\end{equation}
which is larger than the prediction of both the standard QMF theory, 
$\lambda_c^{qmf} = 1/\sqrt{N}$, and the pair approximation developed 
in Ref.~\cite{Cator12}, $\lambda_c\approx  1.37/\sqrt{N}$. 
Our result explains very well the pre-factor larger than 1 observed in
simulations of the SIS model on star graphs~\cite{Ferreira12}.

\begin{figure}[ht]
 \centering
 \includegraphics[width=8cm]{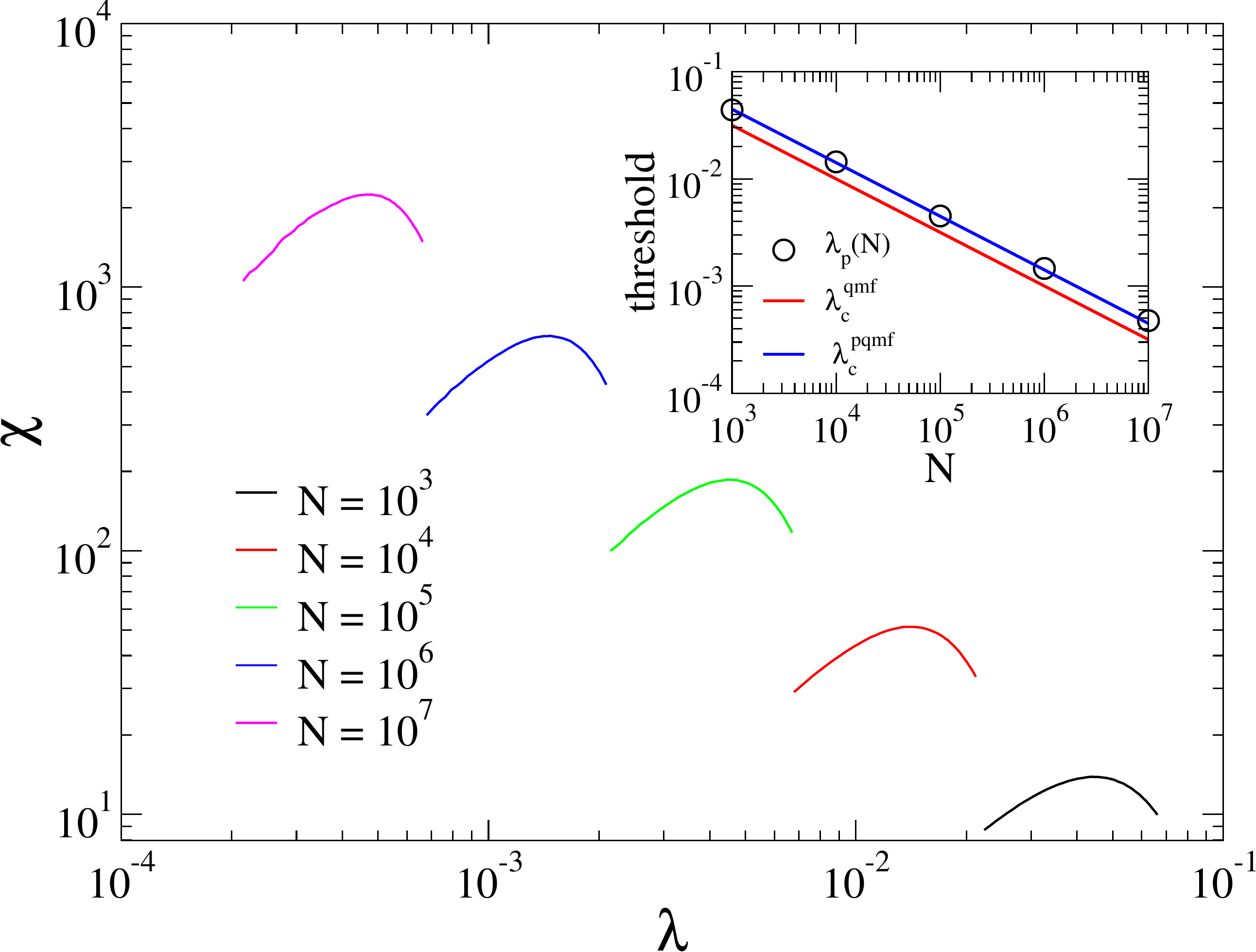}
 \caption{Susceptibility against infection rate for star graphs of
different sizes. Inset compares the threshold predicted by one-vertex and pair QMF theories
with the position of the peaks of the susceptibility.}
 \label{fig:star}
\end{figure}

Figure~\ref{fig:star} shows the susceptibility against the infection rate
exhibiting a rounded peak characteristic of SIS model on star graphs~\cite{Ferreira12}. The
inset of Fig.~\ref{fig:star} compares the thresholds of the  simple and the pair QMF
theories with the positions of the susceptibility peaks for different network
sizes. The last one is assumed as the real threshold of finite networks~\cite{Ferreira12}. 
An excellent agreement between simulations and pair QMF results is obtained
implying a remarkable improvement in relation to the standard QMF approach.
However, it is important to emphasize that the pair QMF is still an approximation and
that a small discrepancy (less than 5\%) is observed for the largest investigated system.

A wheel graph is defined as a regular chain of $N$ vertices and periodic
boundary conditions and a central vertex connected to all vertices of the chain,
Fig.~\ref{fig:simple}(c). This network is interesting in the context of epidemic
spreading due to high clustering coefficient $\lrangle{c}\rightarrow 2/3$ that
contrasts with the null clustering coefficient of the star graph. Following steps
similar to those performed for the star graph, one finds \begin{equation}
\lambda_c = \sqrt{\frac{2}{N}} - \frac{3}{N}+\Oc{(N^{-3/2})}. \end{equation} 
{This threshold is essentially the same obtained for the star graph, which 
was confirmed in simulations (data not shown).
Triangles, characteristic of clustered networks, enhance dynamical correlations
and, therefore, are expected to interfere in the thresholds of dynamical
processes in general~\cite{Rozhnova,ronan}. 
So, the clustering-independence observed for wheel graphs
must be a characteristic  of transitions ruled by the star sub-graph centered at the most
connected vertex~\cite{Castellano12}.}

For arbitrary random networks, the largest eigenvalue of Eq.~(\ref{eq:Lij}) can be
numerically determined~\cite{NR}. We considered the uncorrelated configuration
model (UCM)~\cite{Catanzaro05} with  minimal vertex degree fixed to $k_{min}=3$
and a structural cutoff $k_{c}=N^{1/2}$ in the degree distribution to
investigate SIS dynamics via both Eq.~(\ref{eq:Lij}) and quasi-stationary
simulations. It is known that fluctuations of the degree of the most connected
vertex drastically change the position of the threshold for
$\gamma>3$~\cite{Castellano10}. Therefore, we did our analysis for networks with
$k_{max}\equiv\lrangle{k_{max}}$~\cite{Castellano10,Ferreira12}, in order to
have representative results from a single sample. 

\begin{figure}[ht]
 \centering
 \includegraphics[width=7.5cm]{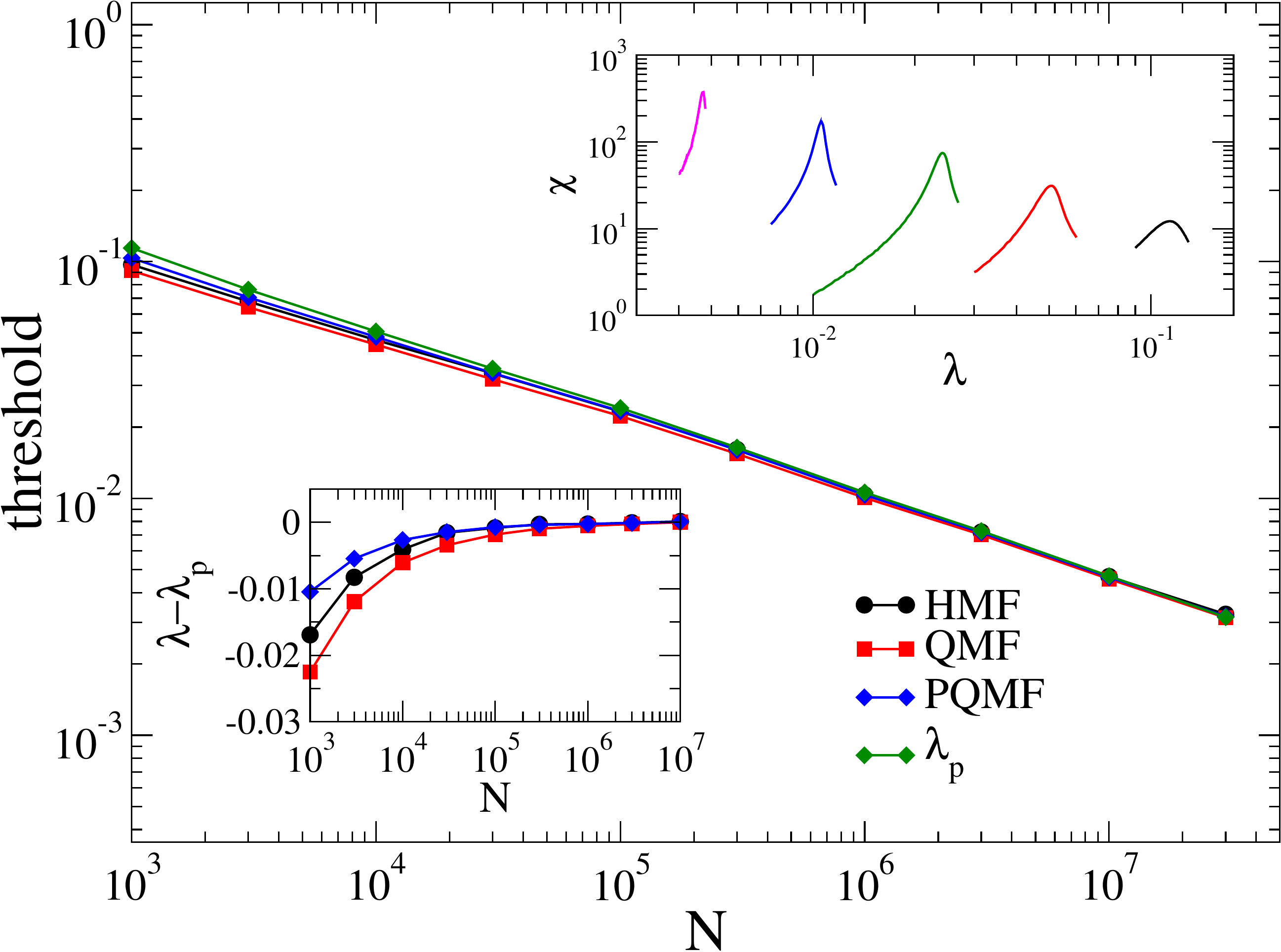}
 \caption{Thresholds against network size for SIS model on UCM networks with
degree exponent $\gamma=2.25$. The top inset shows the susceptibility curves
against infection rate for $N=10^3,10^4,10^5,10^6,$ and $10^7$ (from the right
to the left) used to determine the thresholds in simulations (position of the peaks
$\lambda_p$). The bottom inset shows the difference between
theoretical thresholds and the peaks in the susceptibility curves.}
 \label{fig:lbcg225}
\end{figure}

For the range $2<\gamma<2.50$, the QMF and HMF theories are essentially
equal~\cite{Castellano10} and both theories agree with thresholds estimated from
the peaks of the susceptibility curves~\cite{Ferreira12}. We investigated networks
with $\gamma=2.25$ and verified that the pair QMF is very close to QMF and, consequently,
to HMF theories, being  the pair QMF the one closer to the
simulation results. Figure~\ref{fig:lbcg225} compares the three theories with the
peak of the susceptibility (top inset). One can see that pair QMF theory fits 
better the thresholds for small networks (bottom inset).

\begin{figure}[ht]
 \centering
 \includegraphics[width=7.5cm]{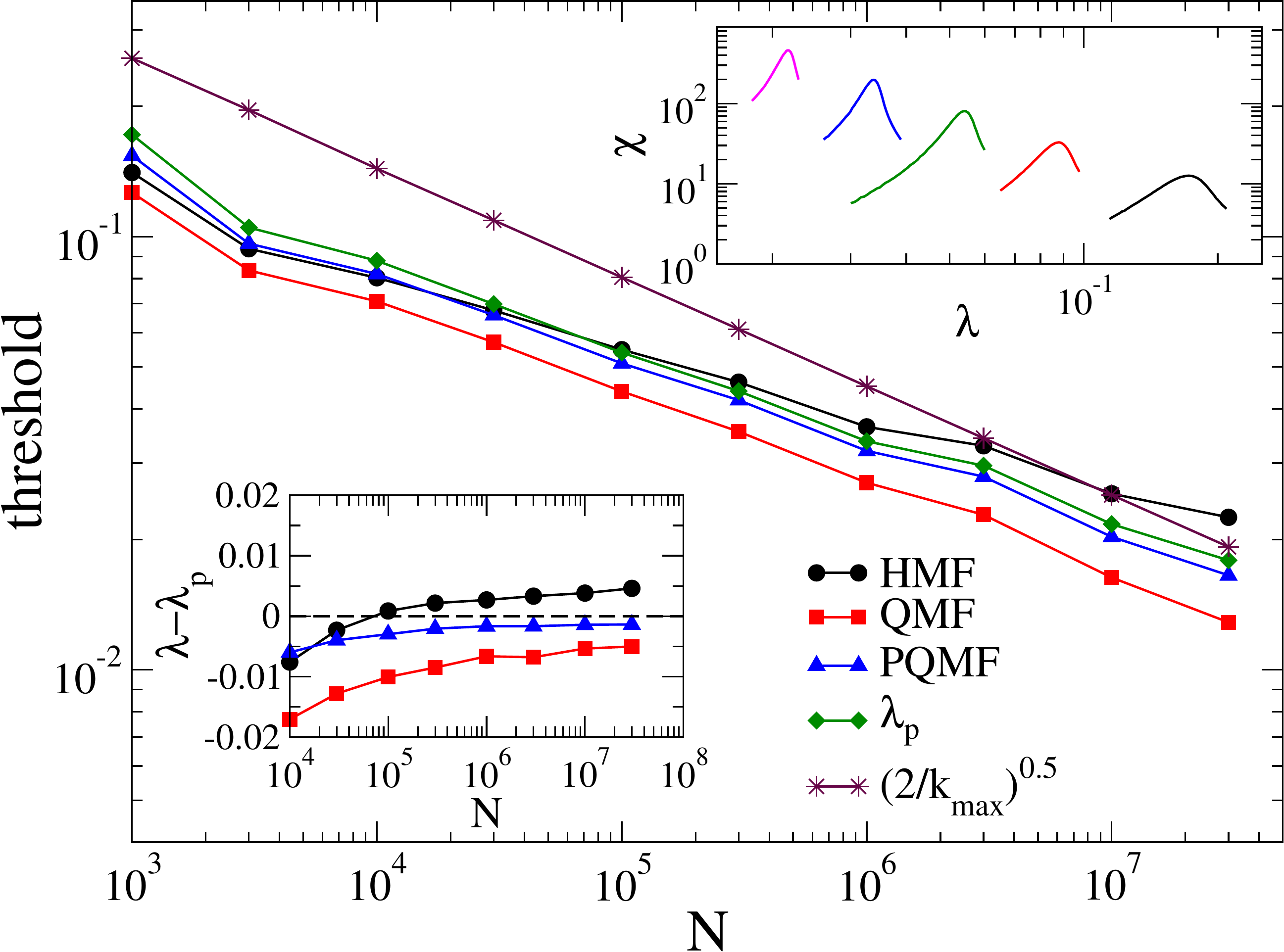}
 \caption{The  same analysis of Fig.~\ref{fig:lbcg225} for $\gamma=2.75$.}
 \label{fig:lbcg275}
\end{figure}

Figure~\ref{fig:lbcg275} shows the thresholds for $\gamma=2.75$. In this case,
the pair QMF consists in a great improvement in relation to the standard QMF (bottom inset of 
Fig.~\ref{fig:lbcg275}). The difference is less than 8\% while standard QMF 
errs in approximately 30\%. It was proposed
that the star sub-graph composed by the most connected vertex and its neighbors
is responsible by sustaining activity in random  network with
$\gamma>2.5$~\cite{Castellano12}. The threshold for the corresponding 
star sub-graph, $\lambda_c^{star}=\sqrt{2/k_{max}}$, is also included in Fig.~\ref{fig:lbcg275}.
One can see that this threshold converges to the pair QMF as network increases 
confirming that the star sub-graph is actually the structure responsible by 
sustaining the epidemics in the entire network.
Notice that HMF theory does not capture
the scaling of threshold against the network size~\cite{Ferreira12}.

\begin{figure}[ht]
 \centering
 \includegraphics[width=8cm]{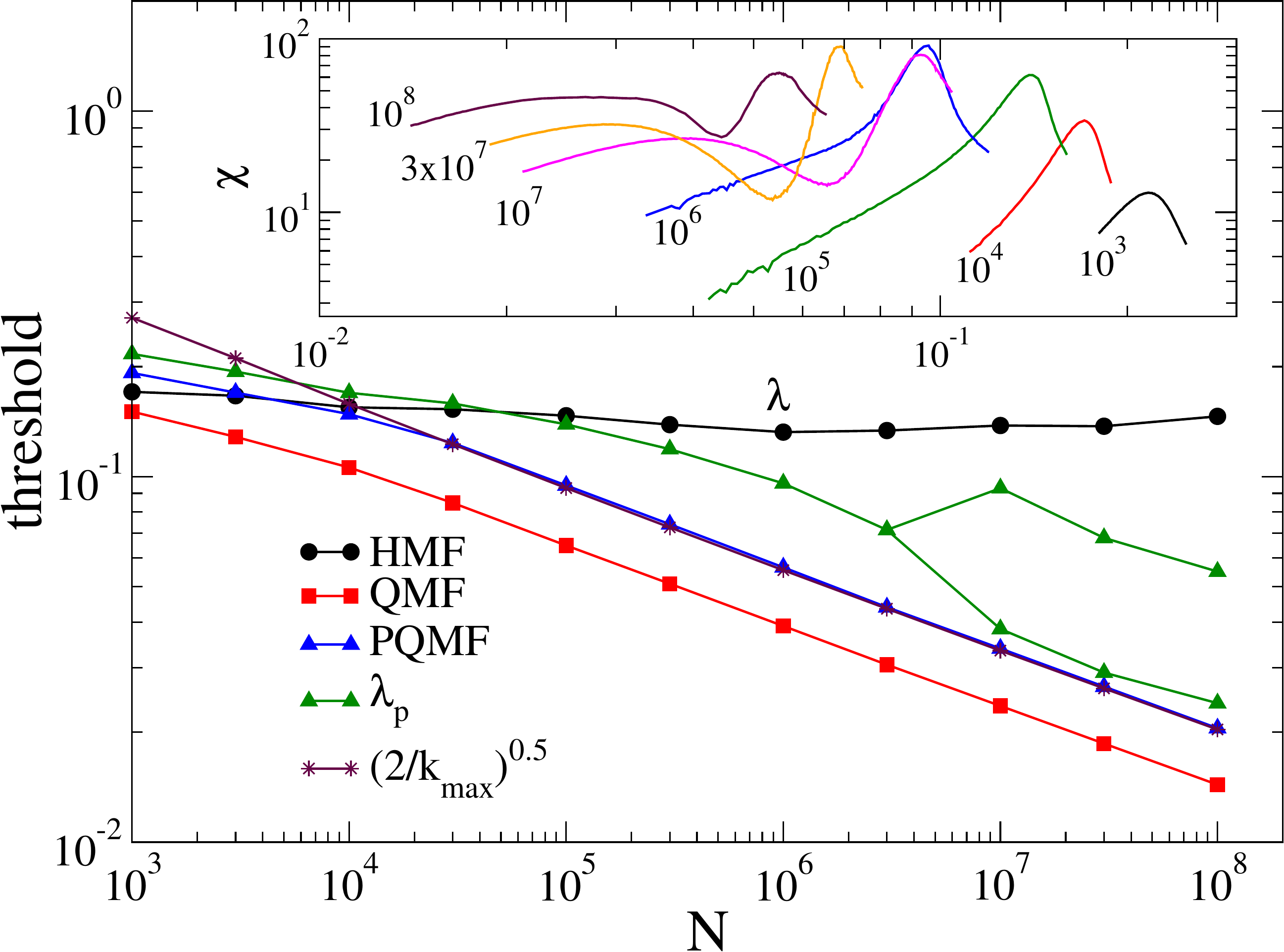}
 \caption{Thresholds against network size for the SIS model on random networks
with degree exponent $\gamma=3.50$. 
Inset shows
the susceptibility curves against infection rate for different sizes (indicated
near each curve) used to determine the thresholds (peaks) in simulations.}
 \label{fig:lbcg350}
\end{figure}

The most drastic differences among theories appear for $\gamma>3$, as
illustrated in Fig.~\ref{fig:lbcg350} where we show the threshold against size
for $\gamma=3.50$. While HMF theory predicts a finite threshold both simple and
pair QMF theories yield asymptotically vanishing thresholds. The $\chi$ \textit{vs}.
$\lambda$ curves have a single sharp peak for small sizes but a secondary
rounded peak at small $\lambda$ emerges for very large sizes. Such doubly peaked
susceptibility was interpreted as the competition between two mechanisms
triggering the epidemic spreading in the network: The activity in the star sub-graph
centered at the most connected vertex against that in the most densely connected component 
of the network~\cite{Ferreira12}. These competing mechanisms are also associated 
to the localization/delocalization of the epidemic activity in networks~\cite{Goltsev12}. 
The peaks at small $\lambda$ are well
fitted by the pair QMF theory (less than 10\% of difference against 40\% for QMF
theory).  The peak at larger $\lambda$ is not captured in our theoretical
approach. Notice that the threshold of the pair QMF theory quickly converges to
that of a star sub-graph $\lambda_c^{star}=\sqrt{2/k_{max}}$. 
{Non-local mean-field
approaches are potential candidates to explain the rightmost 
peaks~\cite{boguna2013nature}}


In conclusion, we have investigated the epidemic spreading of the SIS model
performing a  mean-field pair approximation. Our theory is an extension of the
quenched mean-field (QMF) theory~\cite{Wang03,Castellano10} that includes the
actual network connectivity. 
The dynamical correlations are introduced by means
of a pair approximation~\cite{Avraham92}. Analytical expressions for the epidemic thresholds
are presented for some simple networks while the thresholds for the most
interesting case of random networks with a power law degree distribution
$P(k)\sim k^{-\gamma}$ are obtained from the numerical solution of an eigenvalue
problem.

We compared our pair theory with the one-vertex QMF and
heterogeneous mean-field (HMF) theories and with the peaks of susceptibility as
a function of infection rate obtained in quasi-stationary simulations
~\cite{Ferreira12}.
We have shown that the thresholds in the pair QMF theory as a function of the network size 
scale as in the standard QMF theory. However, the pair QMF
thresholds are quantitatively much closer to the simulation results than those
of standard QMF.

Our theoretical approach represents an important advance in relation to other
improvements of the  QMF theory, which are that limited to small
networks~\cite{Cator12,Gleeson11} whereas we were able to 
investigate networks as large as $N=10^8$. Despite of the considerable 
improvement when compared to QMF theory, our approach is still an approximation. 
Certainly, a $n$-vertex theory with $n>2$ should enhance the accuracy of the 
threshold determination. Also, the critical exponents associated to the transition 
are still an open problem in our approach. The pair QMF theory 
can be extended to other dynamical processes taking place on  the 
complex networks, for which pair approximations have exhibited great improvement
in relation to one-site mean-field 
theories~\cite{FFCR11,sander_phase_2013,Juhasz12,Pugliese09}. 

 This work was partially supported by the Brazilian agencies CNPq and FAPEMIG. 
ASM thanks  the financial support from CAPES. Authors thank the enlightening 
discussions with C. Castellano and R. Pastor-Satorras.


\end{document}